\documentstyle[editedvolume,numreferences,epsf,mathptmx]{crckapb}

\begin{opening}
\title{CURRENT NOISE IN DIFFUSIVE SNS JUNCTIONS IN THE INCOHERENT
MAR REGIME (A REVIEW) }

\author{E.V. Bezuglyi}\author{E.N. Bratus'}
\institute {B. Verkin Institute for Low Temperature Physics and \\
Engineering, 61103 Kharkov, Ukraine }

\author{V.S. Shumeiko} \author{G. Wendin}
\institute {Chalmers University of Technology, S-41296 Gothenburg, \\ Sweden}
\end{opening}

\runningtitle{ CURRENT NOISE IN DIFFUSIVE SNS JUNCTIONS }
\begin{document}

\section{Introduction}
\subsection{Multiple Andreev reflections}

During last decade considerable progress has been made in the investigation
and understanding the mechanisms of current transport in mesoscopic
superconducting junctions. The term mesoscopic here refers to the junctions
where bulk superconducting electrodes in equilibrium (reservoirs) are
connected by small non-superconducting region with the size smaller than any
inelastic mean free path. Such junctions include metallic atomic-size
contacts, tunnel junctions, diffusive (metallic) and ballistic (2D electron
gas) SNS junctions. A common feature of all these structures concerns the
fact that the quasiparticles injected in the junction at zero temperature
cannot escape into the reservoir unless the applied voltage is larger than
the superconducting energy gap in the reservoir, $eV>2\Delta$. In 1963,
Schrieffer and Wilkins suggested that the necessary energy for the
quasiparticle transmission at subgap voltage, $eV<2\Delta$ can be provided by
transferring Cooper pairs between the reservoirs \cite{SW}.

\begin{figure}
\centerline{\epsfxsize=8.5cm \hspace{2.6cm}{\epsffile{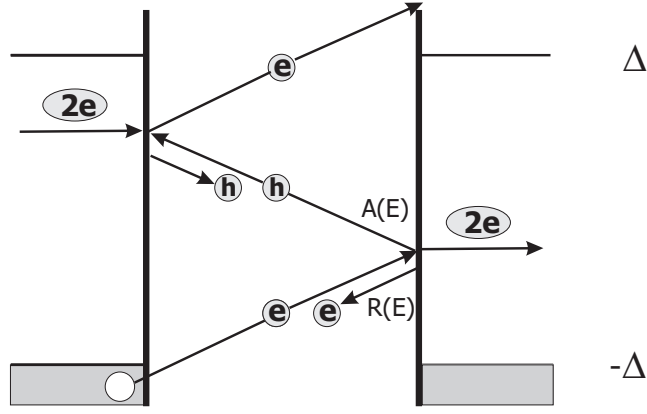}}}
\caption{Multiple Andreev reflections in superconducting junction.
Quasiparticle energy increases by $eV$ after every traversal of the junction,
generating spectral flow across the energy gap, $(-\Delta,\Delta)$. At the
interface,
quasiparticle undergoes Andreev reflection with probability $A(E)$, and
normal reflection with probability $R(E)$.
} \label{MAR}
\end{figure}

The microscopic mechanism for such multiparticle transport, multiple Andreev
reflections (MAR), was suggested in 1982 by Klapwijk, Blonder, and Tinkham
\cite{KBT 82,OTBK}. According to the MAR scenario formulated in terms of the
scattering theory, injected quasiparticles repeatedly undergo Andreev
reflections from the superconducting reservoirs, gaining energy $eV$ during
each traversal of the junction, which allows them to eventually escape from
the junction, see Fig. \ref{MAR}. As the result, a spectral flow across the
energy gap is generated, which creates strongly non-equilibrium quasiparticle
distribution within the contact area. This mechanism explains the nature of
the dissipative current in voltage biased junctions. Also it allows one to
anticipate considerable enhancement of the current shot noise. Indeed,
transmission of one quasiparticle across the energy gap at applied voltage
$V$ requires $N=\mathop{\mbox{Int}}(2\Delta/eV)$ Andreev reflections
[Int($x$) denoting the integer part of $x$], which transfer the total charge,
$q^{\it{eff}}=(N+1)e$, between the electrodes. This enhancement of the
transmitted charge gives rise to the enhancement of the current shot noise
compared to the case of normal junction, according to the Schottky formula,
$S= 2q^{\it{eff}}I$, where $S$ is the spectral density of the noise at zero
frequency.

The mechanism of MAR is general for all kinds of superconducting junctions.
However, the MAR transport regime appears differently in junctions having
small and large lengths. In short junctions where the distance between the
superconducting electrodes is small compared to the superconducting coherence
length, $\xi_0=\hbar v_F/\Delta$, such as point contacts and tunnel
junctions, consequent Andreev reflections are fully coherent. Essential
feature of this coherent MAR regime is the ac Josephson effect, and also
highly non-linear $I$-$V$ characteristics with the subharmonic gap structure
(SGS) -- sequence of current structures at voltages $eV=2\Delta/n$ ($n$ is an
integer) \cite{BdGSIS}.

The same features, the ac Josephson effect and SGS, appear also in ballistic
SNS junctions and short diffusive SNS junctions \cite{Zaitsev 98,Zaitsev 90}.
In diffusive SNS junctions, the electron-hole coherence in the normal metal
persists over a distance of the coherence length $\xi_E = \sqrt{\hbar{\cal
D}/2E}$ from the superconductor (${\cal D}$ is the diffusion constant). The
overlap of coherent proximity regions induced by both SN interfaces creates
an energy gap in the electron spectrum of the normal metal. In short
junctions with a small length $d$ compared to the coherence length,
$d\ll\xi_\Delta$, and with a large proximity gap of the order of the energy
gap $\Delta$ in the superconducting electrodes, the phase coherence covers
the entire normal region.
\begin{figure}
\centerline{\epsfxsize=8.5cm\epsffile{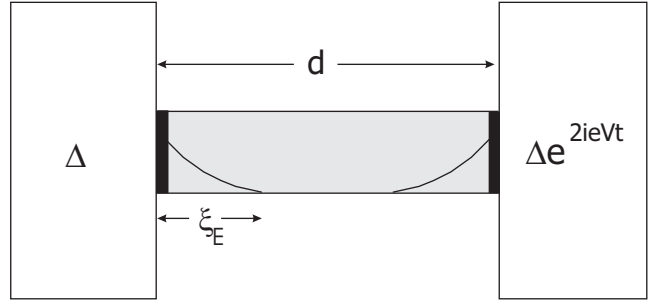}}
\caption{Proximity effect in long diffusive SNS junctions. Shaded region
indicates diffusive metal connecting superconducting reservoirs marked with
$\Delta$; bold lines indicate interface resistances. Superconducting
correlations exponentially decay over length $\xi_E$, which is small compared
to the junction length $d$} \label{SNS}
\end{figure}

Rather different, incoherent MAR regime occurs in long diffusive SNS
junctions with a small proximity gap of the order of Thouless energy $E_{
{\it Th}}={\hbar\cal D}/d^2\ll \Delta$ \cite{Bezuglyi2000}. If $E_{{\it Th}}$
is also small on a scale of applied voltage, $E_{ {\it Th}}\ll eV$, then the
coherence length $\xi_E$ is much smaller than the junction length at all
relevant energies $E\sim \mathop{\mathrm min}(eV,\Delta)$. In this case, the
proximity regions near the SN interfaces become virtually decoupled, as shown
in Fig. \ref{SNS}, and the Josephson current oscillations are strongly
suppressed. At the same time, the quasiparticle distribution inside the
energy gap is strongly non-equilibrium as soon as an inelastic mean free path
exceeds the junction length, $l_\varepsilon \gg d$, because the subgap
electrons must undergo many (incoherent) Andreev reflections before they
enter the reservoir.

It is interesting that in the junctions with transparent interfaces, the
complex transport mechanism of incoherent MAR is not clearly revealed by
$I$-$V$ characteristics of the junction, which are close to the Ohm's law.
Thus the excess shot noise becomes the central characteristic of interest in
this regime. Quasi-linear behavior of $I$-$V$ characteristics in the
incoherent MAR regime can be understood from the following argument. The MAR
transport in real space through the junction with normal resistance $R_N$ is
associated with a spectral current flow along the energy axis through a
series of $N+1$ resistors with the total effective resistance $R = (N+1)R_N$.
Since only electrons incoming within the energy layer $eV$ below the gap
edge, $-\Delta$, participate in the MAR transport, the spectral current $I_p$
is given by the equation $I_p = V/R$. However, each pair of consecutive
Andreev reflections transfers the charge $2e$ through the junction, and the
real current $I$ is therefore $N+1$ times larger than the spectral current:
$I=(N+1)I_p = V/R_N$.
\footnote{In this argument we neglected the small effect of proximity
corrections, which are responsible, together with the interface resistance,
for the SGS in the incoherent MAR regime \cite{Bezuglyi2000}.}
It is clear from this argument, that particular nature of the normal
resistance (e.g. tunnel resistance instead of diffusive metallic resistance)
plays no essential role in the behavior of the current.

\subsection{Current shot noise}

Shot noise in mesoscopic conductors has been extensively studied during
recent years (for a review see \cite{Buttiker2000}). In ballistic normal
electron systems with tunnel barriers, the shot noise is produced by electron
tunnelling through the barriers, and it is described by the Schottky formula
$S = 2eI$ \cite{Jong}. In diffusive metallic wires the shot noise is due to
elastic electron scattering by the impurities, and in this case, an
additional factor $1/3$ appears in the Schottky formula, $S = (2/3)eI$
\cite{1/3Nagaev,1/3theory,1/3exp}. This Fano factor is modified in long wires
whose length exceeds the inelastic scattering length due to the effect of
electron-electron and electron-phonon relaxation
\cite{1/3Nagaev,1/3Nagaeveph,1/3inelastic}.

In normal-superconducting (NS) systems, the current shot noise produced by
the impurity scattering is enhanced at subgap voltages, $eV < \Delta$, by the
factor of two compared to the normal conductors
\cite{doublingtheor,doublingexp}. This is because the subgap current
transport involves one Andreev reflection, which results in the transfer
through the junction of an elementary charge $2e$, instead of $e$. More
elaborated analysis shows further enhancement of the shot noise at small
voltage due to the contribution of the proximity region near the NS interface
\cite{Nazarov}.

In superconducting junctions, the shot noise power is tremendously enhanced
at small voltage by the factor $q^{\it{eff}}/e\sim 2\Delta/eV\gg 1$, because
of the MAR. For the coherent MAR regime this effect has been well
theoretically established \cite{NoiseH,NoiseN,NoiseC} and experimentally
investigated \cite{Klapwijk,Saclay,Hoss}. For the incoherent MAR regime,
similar effect of the shot noise enhancement has been theoretically predicted
in Refs. \cite{Bezuglyi 99,Bezuglyi2001,NagaevSNS}. The experimental
observation of multiply enhanced shot noise in long SNS junctions was
reported in Refs. \cite{Hoss,Strunk,Lefloch}. Although there is the same
mechanism
of the noise enhancement in both cases, there is remarkable difference
between the behavior of the noise power at small voltage in the ballistic and
diffusive junctions. In the ballistic junctions, the current exponentially
decreases with voltage \cite{BdGSIS} while the effective charge grows
inversely proportionally to the voltage, hence the shot noise power also
exponentially decreases. On the other hand, in the diffusive junctions, the
current decay follows a power law. In particular, in the incoherent MAR
regime, the current approximately follows the Ohm's law, and therefore the
noise power approaches a constant value at zero voltage, $S\sim \Delta/R_N$.
This effect, which results from the enhanced effective charge, can be also
explained as the result of strong electron non-equilibrium distribution
developed by the MAR process with an effective noise temperature $T_0\sim
\Delta$.

The finite noise level at zero voltage is a very interesting property of the
incoherent MAR regime, which can be employed for the investigation of the
effect of inelastic relaxation on MAR. The MAR transport regime assumes the
time spent by a quasiparticle within the junction area, $\tau_{{\it dwell}}$,
to be small compared to the inelastic relaxation time $\tau_\varepsilon$.
However, at $eV \rightarrow 0$, the dwelling time infinitely increases
because of MAR, and inelastic relaxation unavoidably starts to play a role.
The inelastic scattering suppresses the spectral flow upwards in energy
generated by MAR, and thus the MAR regime is destroyed. As the result, the
normal region of the junction becomes an equilibrium reservoir, and the SNS
junction turns into two NS junctions connected in series. This is manifested
by the decrease of the noise level at small voltage, the cross over being
controlled by the ratio $\tau_\varepsilon/\tau_{{\it dwell}}$. Precisely this
behavior of the shot noise has been observed in the experiment
\cite{Strunk,Lefloch}.
It is important that because of the absence of the Josephson effect in the
incoherent MAR regime, a small-voltage region of the dissipative current
branch, $E_{Th}\ll eV \ll \Delta$, can be experimentally accessed without
switching to the Josephson branch.

In this paper, we outline a theory of the current shot noise in the
incoherent MAR regime. We will discuss the ``collisionless'' limit of perfect
MAR as well as the effect of inelastic relaxation. While theoretical analysis
of the current shot noise in the coherent MAR regime can be done on the basis
of the scattering theory \cite{NoiseH,NoiseN,Hessling}, the present case
requires different approach, which operates with electron and hole diffusion
flows rather than with ballistic quasiparticle trajectories, and considers
the Andreev reflections as the relationships between these diffusive flows
rather than the scattering events. Such formalism has been developed in Ref.
\cite{Bezuglyi2000}, and it is outlined in the next section.

\section{Circuit theory of incoherent MAR}
\subsection{Microscopic background} 

The system under consideration consists of a normal channel ($0\!<\!x\!<\!d$)
con\-fined between two voltage biased superconducting electrodes, with the
ela\-stic mean free path $\ell$ much shorter than any characteristic size of
the problem. In this limit, the microscopic analysis of current transport can
be performed in the framework of the diffusive equations of nonequilibri\-um
superconductivity \cite{LO} for the $4\!\times\! 4$ matrix Keldysh-Green
function ${\check{G}}(t_1t_2,x)$,
\begin{eqnarray} \label{Keldysh}
&\displaystyle [\check{H},\check{G}]=i \hbar{\cal D}\partial_x \check{J},
\quad \check{J} = \check{G} \partial_x \check{G}, \quad
{\check{G}}^2=\check{1}, \\
\label{Hamiltonian} &\displaystyle
\check{H}=\check{1}[i\hbar\sigma_z\partial_t-e\phi(t) + \hat{\Delta}(t)],
\quad\hat{\Delta}=\Delta e^{i\sigma_z \chi}i\sigma_y,
\end{eqnarray}
where $\Delta,\chi$ are the modulus and the phase of the order parameter, and
$\phi$ is the electric potential. The Pauli matrices $\sigma_i$ operate in
the Nambu space of $2\!\times\!2$ matrices denoted by ``hats'', the products
of two-time functions are interpreted as their time convolutions. The
electric current $I$ per unit area is expressed through the Keldysh component
$\hat{J}^K$ of the matrix current $\check{J}$,
\begin{equation} \label{KeldyshCurrent}
I(t) = (\pi\hbar\sigma_N/4e) \mathop{\mathrm{Tr}} \sigma_z \hat{J}^K (tt,x),
\end{equation}
where $\sigma_N$ is the conductivity of the normal metal.

At the SN interface, the matrix ${\check{G}}$ satisfies the boundary
condition \cite{Kupriyanov 88}
\begin{equation} \label{KeldyshBoundary}
\left(\sigma_N\check{J}\right)_{\pm0}  = \left(2R_{SN}\right)^{-1}
\left[{\check{G}}_{- 0},{\check{G}}_{+0}\right],
\end{equation}
where the indices $\pm 0$ denote the right and left sides of the interface
and $R_{SN}$ is the interface resistance per unit area in the normal state.
Within the model of infinitely narrow potential of the interface barrier,
$U(x)=H\delta(x)$, the interface resistance is related to the barrier
strength $Z=H(\hbar v_F)^{-1}$ as $R_{SN} = 2\ell Z^2 /3\sigma_N$ \cite{BTK}.
It has been shown in Ref.\ \cite{Lambert 97} that Eq.\
(\ref{KeldyshBoundary}) is valid either for a completely transparent
interface ($R_{SN} \rightarrow 0$, ${\check{G}}_{+0}= {\check{G}}_{-0}$) or
for an opaque barrier whose resistance is much larger than the resistance
$R(\ell)=\ell/\sigma_N$ of a metal layer with the thickness formally equal to
$\ell$.

According to the definition of the matrix ${\check{G}}$,
\begin{equation} \label{DefG}
{\check{G}} = \left( \begin{array}{ccc} \hat{g}^R & \hat{G}^K \\ 0  &
\hat{g}^A \end{array} \right), \; \hat{G}^K = \hat{g}^R \hat{f} - \hat{f}
\hat{g}^A,
\end{equation}
Eqs.\ (\ref{Keldysh}) and (\ref{KeldyshBoundary}) represent a compact form of
separate equations for the retarded and advanced Green's functions
$\hat{g}^{R,A}$ and the distribution function $\hat{f}=f_+ +\sigma_z f_-$.
Their time evolution is imposed by the Josephson relation $\chi(t) = 2eVt$
for the phase of the order parameter in the right electrode (we assume $\chi
= 0$ in the left terminal). This implies that the function
$\check{G}(t_1t_2,x)$ consists of a set of harmonics $\check{G}(E_n,E_m,x)$,
$E_n=E+neV$, which interfere in time and produce the ac Josephson current.
However, when the junction length $d$ is much larger than the coherence
length $\xi_E$ at all relevant energies, we may consider coherent
quasiparticle states separately at both sides of the junction, neglecting
their mutual interference and the ac Josephson effect. Thus, the Green's
function in the vicinity of left SN interface can be approximated by the
solution $\hat{g} =\sigma_z \cosh \theta + i\sigma_y \sinh\theta$ of the
static Usadel equations for a semi-infinite SN structure \cite{Zaikin}, with
the spectral angle $\theta (E,x)$ satisfying the equation
\begin{equation} \label{SolutionTheta}
\tanh[\theta(E,x)/ 4] = \tanh[\theta_N(E)/ 4] \exp(-x/\xi_E\sqrt{i}),
\end{equation}
with the boundary condition
\begin{equation} \label{BoundaryTheta}
W\sqrt{i\Delta/E}\sinh (\theta_N-\theta_S) +2\sinh(\theta_N/ 2)=0.
\end{equation}
The indices $S$, $N$ in these equations refer to the superconducting and the
normal side of the interface, respectively.

The dimensionless parameter $W$ in Eq.\ (\ref{BoundaryTheta}),
\begin{equation} \label{W}
W = {R(\xi_\Delta) \over R_{SN}} = {\xi_\Delta \over d r}, \quad  r = {R_{SN}
\over R_N},
\end{equation}
where $R_N=R(d)=d/\sigma_N$ is the resistance of the normal channel per unit
area, has the meaning of an effective barrier transmissivity for the spectral
functions \cite{Bezuglyi-Galaiko 99}. Note that even at large barrier
strength $Z \gg 1$ ensuring the validity of the boundary conditions Eq.\
(\ref{KeldyshBoundary}) \cite{Lambert 97}, the effective transmissivity $W
\sim (\xi_\Delta/\ell)Z^{-2}$ of the barrier in a ``dirty'' system, $\ell \ll
\xi_\Delta$, could be large. In this case, the spectral functions are
virtually insensitive to the presence of a barrier and, therefore, the
boundary conditions Eqs.\ (\ref{KeldyshBoundary}) can be applied to an
arbitrary interface if we approximately consider highly trans\-mis\-sive
interfaces with $W > \xi_\Delta/\ell \gg 1$ as completely transparent, $W =
\infty$.

The distribution functions $f_\pm(E,x)$ are to be considered as global
quantities within the whole normal channel determined by the kinetic
equations
\begin{equation}\label{DiffEq}
\partial_x[D_\pm(E,x) \partial_x f_\pm(E,x)] = 0,
\end{equation}
with dimensionless diffusion coefficients
\begin{eqnarray} \label{DiffTheta}
&\displaystyle D_+ =
(1/4)\mathop{\mathrm{Tr}}\left(1-\hat{g}^R\hat{g}^A\right) = \cos^2
\mathop{\mathrm{Im}}\theta,
\\ \nonumber
&\displaystyle D_- = (1/4)\mathop{\mathrm{Tr}}
\left(1-\sigma_z\hat{g}^R\sigma_z\hat{g}^A\right) = \cosh^2
\mathop{\mathrm{Re}}\theta.
\end{eqnarray}

Assuming the normal conductance of electrodes to be much larger than the
junction conductance, we consider them as equilibrium reservoirs with
unperturbed spectrum, $\theta_S = \mathop{\mathrm{Arctanh}} (\Delta/E)$, and
equilibrium quasiparticle distribution, $\hat{f}_S(E) = f_0(E) \equiv
\tanh(E/2T)$. Within this approximation, the boundary conditions for the
distribution functions at $x = 0$ read
\begin{equation} \label{BoundaryF}
\sigma_N D_+\partial_x f_+(E,0) = G_+(E)[f_+(E,0)-f_0(E)],
\end{equation}
\begin{equation} \label{BoundaryFz}
\sigma_N D_-\partial_x f_-(E,0)=G_-(E)f_-(E,0),
\end{equation}
where
\begin{equation} \label{AB}
G_\pm(E) = R_{SN}^{-1}\left(N_S N_N \mp M_S^\pm M_N^\pm\right),
\end{equation}
\begin{equation} \label{NM}
N(E)= \mathop{\mathrm{Re}}(\cosh\theta), \;M^+(E)+iM^-(E)=\sinh\theta.
\end{equation}
At large energies, $|E| \gg \Delta$, when the normalized density of states
$N(E)$ approaches unity and the condensate spectral functions $M^\pm(E)$ turn
to zero at both sides of the interface, the conductances $G_\pm(E)$ coincide
with the normal barrier conductance; within the subgap region $|E| < \Delta$,
$G_+(E) = 0$.

Similar considerations are valid for the right NS interface, if we eliminate
the time dependence of the order parameter in Eq.\ (\ref{Keldysh}), along
with the potential of right electrode, by means of a gauge transformation
\cite{Artemenko 79}
\begin{equation} \label{Gauge}
{\check{G}}(t_1t_2,x) = \exp(i\sigma_z eVt_1) \widetilde
{{\check{G}}}(t_1t_2,x) \exp(-i\sigma_z eVt_2).
\end{equation}
As a result, we arrive at the same static equations and boundary conditions,
Eqs.\ (\ref{SolutionTheta})-(\ref{NM}), with $x \rightarrow d-x$, for the
gauge-transformed functions $\widetilde{\hat{g}}(E,x)$ and
$\widetilde{\hat{f}}(E,x)$. Thus, to obtain a complete solution, e.g. for the
distribution function $f_-$, which determines the dissipative current,
\begin{equation} \label{Disscurrent}
I = {\sigma_N\over 2e}\int_{-\infty}^\infty dE\, D_- \partial_x f_-,
\end{equation}
one must solve the boundary problem for ${\hat{f}}(E,x)$ at the left SN
interface, and a similar boundary problem for $\widetilde{\hat{f}}(E,x)$ at
the right interface, and then match the distribution function asymptotics
deep inside the normal region by making use of the relationship following
from Eqs.\ (\ref{DefG}), (\ref{Gauge}),
\begin{equation}  \label{GaugeF}
{\hat{f}}(E,x) = \widetilde{\hat{f}}(E +\sigma_z eV,x).
\end{equation}
%

\subsection{Circuit representation of boundary conditions}

In order for this kinetic scheme to conform to the conventional physical
interpretation of Andreev reflection in terms of electrons and holes, we
introduce the following parametrization of the matrix distribution function,
\footnote{The absence of the factor $2$ in an analogous formula (18) in the
paper \cite{Bezuglyi2000} is a typo.}
\begin{equation} \label{Defn}
\hat{f}(E,x)= 1-2\left(\begin{array}{ccc} n^e(E,x) & 0 \\ 0 & n^h(E,x)
\end{array} \right),
\end{equation}
where $n^e$ and $n^h$ will be considered as the electron and hole population
numbers. Deep inside the normal metal region, they acquire rigorous meaning
of distribution functions of electrons and holes, and approach the Fermi
distribution in equilibrium. In this representation, Eqs.\ (\ref{DiffEq})
take the form
\begin{equation} \label{Diffn}
D_\pm(E,x)\partial_x n_\pm(E,x) = \mbox{const} \equiv -I_\pm(E)/\sigma_N,
\end{equation}
where $n_{\pm} = n^e\pm n^h$, and they may be interpreted as conservation
equations for the (specifically normalized) net probability current $I_+$ of
electrons and holes, and for the electron-hole imbalance current $I_-$.
Furthermore, the probability currents of electrons and holes, defined as
$I^{e,h} = (1/2)(I_+ \pm I_-)$, separately obey the conservation equations.
The probability currents $I^{e,h}$ are naturally related to the electron and
hole diffusion flows, $I^{e,h} = -\sigma_N \partial_x n^{e,h}$, at large
distances $x \gg \xi_E$ from the SN boundary. Within the proximity region, $x
< \xi_E$, each current $I^{e,h}$ generally consists of a combination of both
the electron and hole diffusion flows, which reflects coherent mixing of
normal electron and hole states in this region,
\begin{equation}
I^{e,h} = -(\sigma_N / 2)\left[ (D_+\pm D_-)\partial_x n^e + (D_+\mp
D_-)\partial_x n^h \right].
\end{equation}
\begin{figure}
\epsfxsize=8cm\centerline{\epsffile{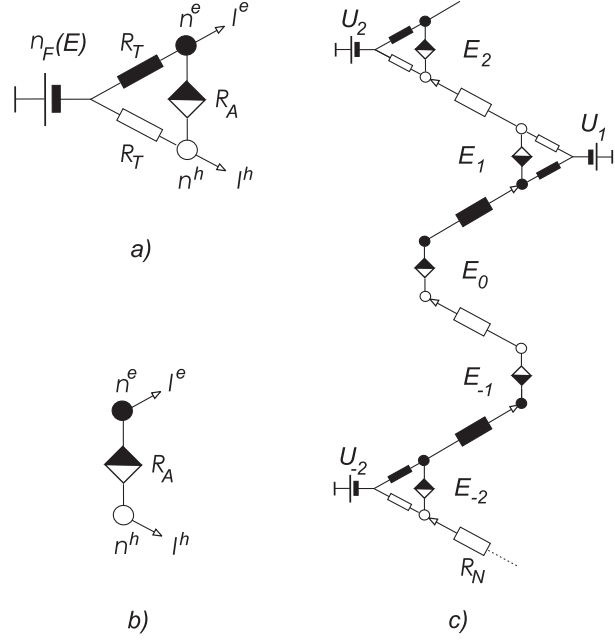}}
\caption{Elementary equivalent circuits representing boundary conditions in
Eq.\ (\ref{BoundaryOhm}) for the electron and hole population numbers
$n^{e,h}(E,0)$ and probability currents $I^{e,h}(E)$, at energies outside the
gap, $|E| > \Delta$ (a), and within the subgap region, $|E| < \Delta$ (b);
equivalent network in ($x,E$)-space for incoherent MAR in SNS junction (c).
Filled and empty symbols stand for electron- and hole- related elements,
respectively; half-filled squares denote Andreev resistors; $U_n =
n_F(E_n)$.} \label{SpaceCircuit}
\end{figure}

In terms of electrons and holes, the boundary conditions in Eqs.\
(\ref{BoundaryF}), (\ref{BoundaryFz}) read
\begin{equation} \label{BoundaryOhm}
I^{e,h} = G_T\left(n_F - n^{e,h}\right) \mp G_A\left(n^e - n^h\right),
\end{equation}
where
\begin{equation} \label{DefCond}
G_T = G_+, \; G_A = (G_- -G_+)/2.
\end{equation}

Each of the equations (\ref{BoundaryOhm}) may be clearly interpreted as a
Kirchhoff rule for the electron or hole probability current flowing through
the effective circuit (tripole) shown in Fig.\ \ref{SpaceCircuit}(a). Within
such interpretation, the nonequilibrium populations of electrons and holes
$n^{e,h}$ at the interface correspond to ``potentials'' of nodes attached to
the ``voltage source'' --  the Fermi distribution $n_F(E)$ in the
superconducting reservoir -- by ``tunnel resistors'' $R_T(E) = G_T^{- 1}(E)$.
The ``Andreev resistor'' $R_A(E) = G_A^{-1}(E)$ between the nodes provides
electron-hole conversion (Andreev reflection) at the SN interface.

The circuit representation of the diffusive SN interface is analogous to the
scattering description of ballistic SN interfaces: the tunnel and Andreev
resistances play the same role as the normal and Andreev reflection
coefficients in the ballistic case \cite{BTK}. For instance, at $|E|>\Delta$
[Fig.\ \ref{SpaceCircuit}(a)], the probability current $I^e$ is contributed
by equilibrium electrons incoming from the superconductor through the tunnel
resistor $R_T$, and also by the current flowing through the Andreev resistor
$R_A$ as the result of hole-electron conversion. Within the subgap region,
$|E|<\Delta$ [Fig.\ \ref{SpaceCircuit}(b)], the quasiparticles cannot
penetrate into the superconductor, $R_T = \infty$, and the voltage source is
disconnected, which results in detailed balance between the electron and hole
probability currents, $I^e = -I^h$ (complete reflection). For the perfect
interface, $R_A$ turns to zero, and the electron and hole population numbers
become equal, $n^e=n^h$ (complete Andreev reflection). The nonzero value of
the Andreev resistance for $R_{SN} \neq 0$ accounts for suppression of
Andreev reflection due to the normal reflection by the interface.

Detailed information about the boundary resistances can be obtained from
their asymptotic expressions in Ref.\ \cite{Bezuglyi2000}. In particular,
$R_\pm(E)$ turns to zero at the gap edges due to the singularity in the
density of states which enhances the tunnelling probability. The resistance
$R_-(E)$ approaches the normal value $R_{SN}$ at $E \rightarrow 0$ due to the
enhancement of the Andreev reflection at small energies, which results from
multiple coherent backscattering of quasiparticles by the impurities within
the proximity region.
\footnote{This property is the reason for the re-entrant behavior of the
conductance of high-resistive SIN systems \cite{Volkov-Klapwijk 92,VZK} at
low voltages. In the MAR regime, one cannot expect any reentrance since
quasiparticles at all subgap energies participate in the charge transport
even at small applied voltage.}
\begin{figure}
\epsfxsize=7.5cm\centerline{\epsffile{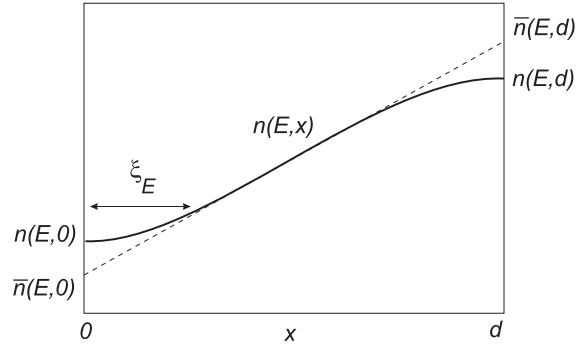}}
\caption{Qualitative behavior of population numbers within the normal channel
(solid curve). The edge distortions of the linear $x$-dependence of
population numbers, Eq.\ (\ref{Asympt}), occur within the proximity regions.
The difference between the boundary population numbers $n(E,0)$, $n(E,d)$ and
their effective values $\overline n(E,0)$, $\overline n(E,d)$ for true normal
electrons and holes is included in the renormalization of the boundary
resistances, Eq.\ (\ref{RenormAB}).} \label{SketchN}
\end{figure}

The proximity effect can be incorporated into the circuit scheme by the
following way. We note that the diffusion coefficients $D_\pm$ in Eq.\
(\ref{DiffTheta}) turn to unity far from the SN boundary, and therefore the
population numbers $n^{e,h}$ become linear functions of $x$,
\begin{equation} \label{Asympt}
n^{e,h}(E,x) \approx {\overline n}^{e,h}(E,0) - R_N I^{e,h}(E)x/d.
\end{equation}

This equation defines the renormalized population numbers ${\overline
n}^{e,h}(E,0)$ at the NS interface, which differ from $n^{e,h}(E,0)$ due to
the proximity effect, as shown in Fig.\ \ref{SketchN}. These quantities have
the meaning of the true electron/hole populations which would appear at the
NS interface if the proximity effect had been switched off. It is possible to
formulate the boundary conditions in Eq.\ (\ref{BoundaryOhm}) in terms of
these population numbers by including the proximity effect into
renormalization of the tunnel and Andreev resistances. To this end, we will
associate the node potentials with renormalized boundary values
$\overline{n}^{e,h}(E,0) = (1/2)[\overline{n}_+(E,0) \pm
\overline{n}_-(E,0)]$ of the population numbers, where ${\overline
n}_\pm(E,0)$ are found from the exact solutions of Eqs.\ (\ref{Diffn}),
\begin{equation} \label{Renormn}
{\overline n}_\pm(E,0) =  n_\pm(E,0) - m_\pm(E)I_\pm(E).
\end{equation}
Here $m_\pm(E)$ are the proximity corrections to the normal metal resistance
at given energy for the probability and imbalance currents, respectively,
\begin{equation}
m_\pm(E)=\pm R_N  \int_0^\infty {dx\over d} \left| D_\pm^{-1}(E,x)-1\right|.
\end{equation}
It follows from Eq.\ (\ref{Renormn}) that the same Kirchhoff rules as in
Eqs.\ (\ref{BoundaryOhm}), (\ref{DefCond}) hold for ${\overline
n}^{e,h}(E,0)$ and $I^{e,h}(E)$, if the bare resistances $R_\pm$ are
substituted by the renormalized ones,
\begin{equation} \label{RenormAB}
R_\pm(E) \rightarrow \overline{R}_\pm(E) = R_\pm(E)+ m_\pm(E).
\end{equation}

In certain cases, there is an essential difference between the bare and
renormalized resistances, which leads to qualitatively different properties
of the SN interface for normal electrons and holes compared to the properties
of the bare boundary. Let us first discuss a perfect SN interface with
$R_{SN} \rightarrow 0$. Within the subgap region $|E| < \Delta$, the bare
tunnel resistance $R_T$ is infinite whereas the bare Andreev resistance $R_A$
turns to zero; this corresponds to complete Andreev reflection, as already
explained. However, the Andreev resistance for normal electrons and holes,
$\overline R_A(E) = 2m_-(E)$, is finite and negative,
\footnote{In the terms of the circuit theory, this means that the ``voltage
drop'' between the electron and hole nodes is directed against the
probability current flowing through the Andreev resistor.}
which leads to enhancement of the normal metal conductivity within the
proximity region \cite{VZK,Stoof}. At $|E|>\Delta$, the bare tunnel
resistance $R_T$ is zero, while the renormalized tunnel resistance $\overline
R_T(E) = m_+(E)$ is finite (though rapidly decreasing at large energies).
This leads to suppression of the probability currents of normal electrons and
holes within the proximity region, which is to be attributed to the
appearance of Andreev reflection. Such a suppression is a global property of
the proximity region in the presence of sharp spatial variation of the order
parameter, and it is similar to the over-the-barrier Andreev reflection in
the ballistic systems. In the presence of normal scattering at the SN
interface, the overall picture depends on the interplay between the bare
interface resistances $R_\pm$ and the proximity corrections $m_\pm$; for
example, the renormalized tunnel resistance $\overline R_T(E)$ diverges at
$|E|\rightarrow \Delta$, along with the proximity correction $m_+(E)$, in
contrast to the bare tunnel resistance $R_T(E)$. This indicates complete
Andreev reflection at the gap edge independently of the transparency of the
barrier, which is similar to the situation in the ballistic systems where the
probability of Andreev reflection at $|E|=\Delta$ is always equal to unity.

\subsection{MAR networks} 

To complete the definition of an equivalent MAR network, we have to construct
a similar tripole for the right NS interface and to connect boundary values
of population numbers (node potentials) using the matching condition in Eq.\
(\ref{GaugeF}) expressed in terms of electrons and holes,
\begin{equation} \label{GaugeN}
n^{e,h}(E,x) = \widetilde{n}^{e,h}(E\pm eV,x).
\end{equation}

Since the gauge-transformed distribution functions $\widetilde{f}_\pm$ obey
the same equations Eq.\ (\ref{DiffEq})-(\ref{NM}), the results of the
previous Section can be applied to the functions $\widetilde{n}^{e,h}(E)$ and
$-\widetilde{I}^{e,h}(E)$ (the minus sign implies that $\widetilde{I}$ is
associated with the current incoming to the right-boundary tripole). In
particular, the asymptotics of the gauge-transformed population numbers far
from the right interface are given by the equation
\begin{equation} \label{AsymptGauge}
\widetilde{n}^{e,h}(E,x) \approx \widetilde{\overline{n}}^{e,h}(E,d) + R_N
\widetilde{I}^{e,h}(E)\left(1-x/d\right).
\end{equation}

After matching the asymptotics in Eqs. (\ref{Asympt}) and (\ref{AsymptGauge})
by means of Eq.\ (\ref{GaugeN}), we find the following relations,
\begin{equation} \label{GaugeI}
I^{e,h}(E) = \widetilde{I}^{e,h}(E\pm eV),
\end{equation}
\begin{equation} \label{Matching}
\overline{n}^{e,h}(E,0) - \widetilde{\overline{n}}^{e,h}(E\pm eV,d) = R_N
I^{e,h}(E).
\end{equation}
From the viewpoint of the circuit theory, Eq.\ (\ref{Matching}) may be
interpreted as Ohm's law for the resistors $R_N$ which connect energy-shifted
boundary tripoles, separately for the electrons and holes, as shown in Fig.\
\ref{SpaceCircuit}(c).

The final step which essentially simplifies the analysis of the MAR network,
is based on the following observation. The spectral probability currents
$I^{e,h}$ yield opposite contributions to the electric current in Eq.\
(\ref{Disscurrent}),
\begin{equation} \label{CurrentIeh}
I = {1\over 2e}\int_{-\infty}^{\infty} dE \left[I^e(E)-I^h(E)\right],
\end{equation}
\begin{figure}
\epsfxsize=8.5cm\centerline{\epsffile{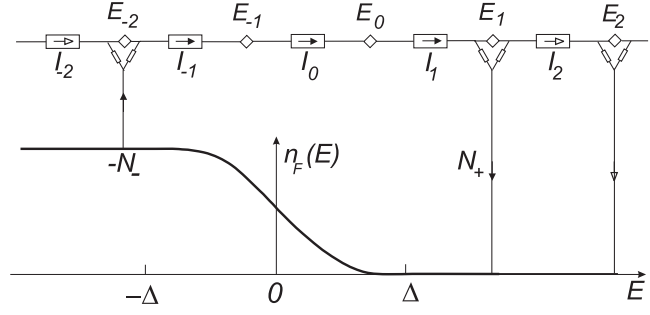}}
\caption{MAR network of Fig.\ \ref{SpaceCircuit}(c) in energy space. The
nodes outside the gap are connected with the distributed voltage source
$n_F(E)$ (bold curve); the subgap nodes are disconnected from the voltage
source.} \label{EnergyCircuit}
\end{figure}
due to the opposite charge of electrons and holes. At the same time, these
currents, referred to the energy axis, transfer the charge in the same
direction, viz., from bottom to top of Fig.\ \ref{SpaceCircuit}(c), according
to our choice of positive $eV$. Thus, by introducing the notation $I_n(E)$
for an electric current entering the node $n$, as shown by arrows in Fig.\
\ref{SpaceCircuit}(c),
\begin{equation} \label{DefElectric}
I_n(E) = \left\{ \begin{array}{ccc} I^e(E_{n-1}), & n=2k+1, \\ -I^h(E_n), &
n=2k,
\end{array} \right., \quad E_n = E +neV,
\end{equation}
we arrive at an ``electrical engineering'' problem of current distribution in
an equivalent network in energy space plotted in Fig.\ \ref{EnergyCircuit},
where the difference between electrons and holes becomes unessential. The
bold curve in Fig.\ \ref{EnergyCircuit} represents a distributed voltage
source -- the Fermi distribution $n_F(E)$ connected periodically with the
network nodes. Within the gap, $|E_n|<\Delta$, the nodes are disconnected
from the Fermi reservoir and therefore all partial currents associated with
the subgap nodes are equal.

Since all resistances and potentials of this network depend on $E_n=E+neV$,
the partial currents obey the relationship $I_n(E) = I_k[E+(n-k)eV]$ which
allows us to express the physical electric current, Eq.\ (\ref{CurrentIeh}),
through the sum of all partial currents $I_n$ flowing through the normal
resistors $R_N$, integrated over an elementary energy interval $0<E<eV$,
\begin{equation} \label{Electric}
I = {1\over 2e}\int_{-\infty}^\infty\!\!\! dE\, [I_1(E)+I_0(E)] = {1\over
e}\int_0^{eV}\!\!\!dE\, J(E), \quad J(E) =\!\!\! \sum_{n=-\infty}^{+\infty}
\!\!\!I_n(E).
\end{equation}
The spectral density $J(E)$ is periodic in $E$ with the period $eV$ and
symmetric in $E$, $J(-E)=J(E)$, which follows from the symmetry of all
resistances with respect to $E$.

As soon as the partial currents are found, the population numbers can be
recovered by virtue of Eqs.\ (\ref{Diffn}), (\ref{BoundaryOhm}),
(\ref{Asympt}), and (\ref{DefElectric}),
\begin{equation} \label{Recovern}
n^{e,h}(E,x) = \overline{n}^{e,h}(E,0) \mp R_N I_{1,0}(E)x/ d,
\end{equation}
\begin{equation} \label{Recovern0}
\overline{n}^{e,h}(E,0) =n_F - (1/ 2) \left[ \overline{R}_+(I_1-I_0) \pm
\overline{R}_-(I_1+I_0) \right]
\end{equation}
at $|E|>\Delta$. Within the subgap region, Eq.\ (\ref{Recovern0}) is
inapplicable due to the indeterminacy of product $\overline{R}_+ (I_1- I_0)$.
In this case, one may consider the subgap part of the network as a voltage
divider between the nodes nearest to the gap edges, having the numbers
$-N_-$, $N_+$, respectively, where
\begin{equation} \label{Npm}
N_\pm = \mbox{Int}[(\Delta \mp E)/eV]+1.
\end{equation}
Then the boundary populations at $|E|<\Delta$ become
\begin{equation} \label{SubgapN}
\overline{n}^{e,h}(E,0) = n^{L,R}(E_{\pm N_\pm}) \pm I_0 \left[ N_\pm R_N +
\sum_{k=1}^{N_\pm-1} R_A(E_{\pm k}) \right],
\end{equation}
where $R,L$ indicate the right (left) node of the tripole, irrespectively of
whether it relates to the left (even $n$) or right (odd $n$) interface. The
physical meaning of $n^{R,L}(E_n)$, however, depends on the parity of $n$,
\begin{equation} \label{Nparity}
n^{R,L}(E_n) = \left\{ \begin{array}{ccc}
\overline{n}^{e,h}(E_n,0), & n=2k, \\
\widetilde{\overline{n}}^{h,e}(E_n,d),  & n=2k+1. \end{array}\right.
\end{equation}
The values $n^{R,L}$ in Eq.\ (\ref{SubgapN}) can be found from Eq.\
(\ref{Recovern0}) which is generalized for any tripole of the network in
Fig.\ \ref{EnergyCircuit} outside the gap as
\begin{equation} \label{Recover}
n^{R,L}(E_n) =n_F(E_n) - {1\over 2} \left[ \overline{R}_+(E_n)(I_{n+1}-I_n)
\pm \overline{R}_-(E_n)(I_{n+1}+I_n) \right].
\end{equation}

\begin{figure}
\epsfxsize=7.5cm\centerline{\epsffile{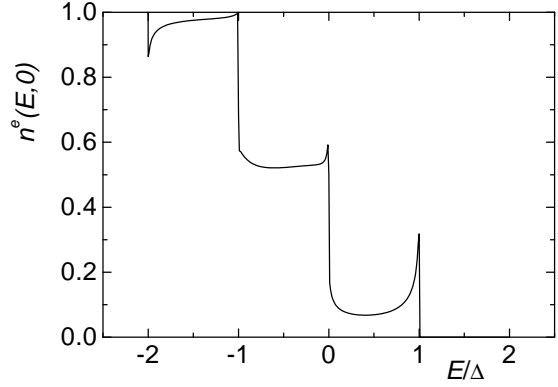}}
\caption{Energy dependence of the electron population number $n^{e}(E,0)$ at
the left interface of the SNS junction with $R_{SN} = R_N$ and $d =
5\xi_\Delta$, at $V = \Delta/e$ and $T=0$.} \label{n(E)}
\end{figure}

As follows from Eqs.\ (\ref{Recovern0}), (\ref{SubgapN}), the energy
distribution of quasiparticles has a step-like form (Fig.\ \ref{n(E)}), which
is qualitatively similar to, but quantitatively different from that found in
OTBK theory \cite{OTBK}. The number of steps increases at low voltage, and
the shape of the distribution function becomes resemblant to a ``hot
electron'' distribution with the effective temperature of the order of
$\Delta$. This distribution is modulated due to the discrete nature of the
heating mechanism of MAR, which transfers the energy from an external voltage
source to the quasiparticles by energy quanta $eV$.

The circuit formalism can be simply generalized to the case of different
transparencies of NS interfaces, as well as to different values of $\Delta$
in the electrodes. In this case, the network resistances become dependent not
only on $E_n$ but also on the parity of $n$. As a result, the periodicity of
the current spectral density doubles: $J(E)=J(E+2eV)$, and, therefore, $J(E)$
is to be integrated in Eq.\ (\ref{Electric}) over the period $2eV$, with an
additional factor $1/2$.

\section{Current shot noise in the MAR regime}

\subsection{Noise of normal conductor}

The main source of the shot noise in long diffusive SNS junctions with
transparent interfaces ($R_{NS}\ll R_N$) is the impurity scattering of
electrons in the normal region of the junction. In this section we apply
the circuit theory to calculate the noise of the normal conductor in the
junction \cite{Bezuglyi2001}. The effect of proximity
regions near NS interfaces can be neglected for $\xi_\Delta\ll d$ since their
length is small compared to the length of the diffusive conductor. The noise
of the diffusive normal region can be calculated within a Langevin approach
\cite{Langevin}.  Following Ref.\ \cite{NS}, in which the Langevin equation
was applied to the current fluctuations in a diffusive NS junction, we derive
an expression for the current noise spectral density in SNS junctions at zero
frequency in terms of the nonequilibrium population numbers $n^{e,h}(E,x)$ of
electrons and holes within the normal metal, $0<x<d$,
\begin{equation} \label{noise_eh}
S = {2\over R_N} \int_0^d {dx \over d}\int_{-\infty}^\infty dE
\left[n^e\left(1-n^e\right) + n^h\left(1-n^h\right)\right].
\end{equation}
The electric current through the junction, Eq. (\ref{Disscurrent}), is given
by
\begin{equation} \label{I1}
I ={d\over 2eR_N} \int_{-\infty}^\infty dE\; \partial_x \left(n^e -
n^h\right).
\end{equation}
\begin{figure}
\epsfxsize=8.5cm\centerline{\epsffile{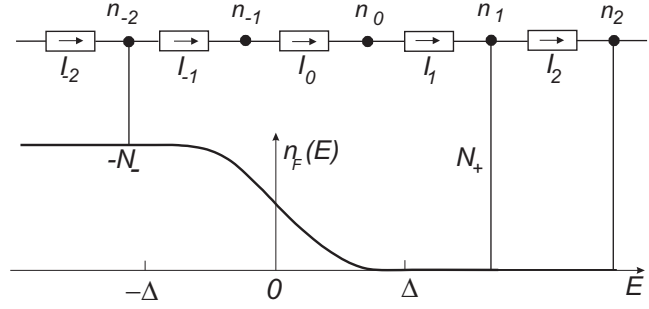}}
\caption{Equivalent MAR network in energy space in the limit of negligibly
small normal reflection and proximity regions at the NS interfaces.}
\label{network}
\end{figure}

The MAR network for the present case, in which the contributions of the
proximity effect and normal scattering at the interfaces have been neglected,
is shown in Fig.\ \ref{network}. Within such model i) the renormalized
resistances coincide with the bare ones, ii) the Andreev reflection inside
the gap is complete: $R_A = 0$, $G_T = 0$ at $|E| < \Delta$, and iii) the
over-the-barrier Andreev reflection, as well as the normal reflection, are
excluded:  $G_A = 0$, $R_T = 0$ at $|E| > \Delta$. As the result, the network
is essentially simplified and represents a series of Drude resistances
connected periodically, at the energies $E_k = E + keV$, with the distributed
``voltage source'' $n_F(E)$. The ``potentials'' $n_k$ of the network nodes
with even numbers $k$ represent equal electron and hole populations
$n^{e,h}_0(E_k)$ at the left NS interface, whereas the potentials of the odd
nodes describe equal boundary populations $n^{e,h}_d(E_k\mp eV)$ at the right
interface. The ``currents'' $I_k$ entering $k$-th node are related to the
probability currents $n^{\prime}(E_k) = \partial n(E_k)/\partial x$ as
$I_k(E) = -\sigma_N n^{e\prime}(E_{k-1})$ (odd $k$) and $I_k(E) = \sigma_N
n^{h\prime}(E_{k})$ (even $k$), and represent partial electric currents
transferred by the electrons and holes across the junction, obeying Ohm's law
in energy space, $I_k = (n_{k-1} - n_k)/R_N$. Within the gap, $|E_k| <
\Delta$, i.e., at $-N_- < k < N_+$, the nodes are disconnected from the
reservoir due to complete Andreev reflection and therefore all currents
flowing through the subgap nodes are equal.

The analytical equations corresponding to the MAR network are as follows: At
$|E| > \Delta$, the bo\-un\-da\-ry populations $n_{0,d}(E) =
n(E,x)\left|_{x=0,d}\right.$ are local-equi\-lib\-rium Fermi functions,
$n^{e,h}_0(E) = n_F(E)$, $n^{e,h}_d(E) = n_F(E\pm eV)$ (we use the potential
of the left electrode as the energy reference level). At subgap energies,
$|E| < \Delta$, the boundary conditions are modified in accordance with the
mechanics of complete Andreev reflection which equalizes the electron and
hole population numbers at a given electrochemical potential and blocks the
net probability current through the NS interface \cite{Bezuglyi2000},
\begin{eqnarray}\label{boundary}
&\displaystyle n^e_0(E) = n^h_0(E), \; n^e_d(E-eV) = n^h_d(E+eV),
\nonumber \\
&\displaystyle n^{e\prime}_0(E) + n^{h\prime}_0(E)=0, \; n^{e\prime}_d(E-eV)
+ n^{h\prime}_d(E+eV) = 0,
\end{eqnarray}
where $n^{\prime}_{0,d}$ are the boundary values of the electron and hole
probability flows $\partial n / \partial x$. The recurrences for boundary
populations and diffusive flows within the subgap region read
\begin{eqnarray} \label{recurr}
&n^{e,h}_0(E-eV) - n^{e,h}_0(E+eV) = \mp 2d n^{e,h\prime}(E\mp eV),
\\ \nonumber
&n^{e,h\prime}_0(E-eV) = n^{e,h\prime}_0(E+eV).
\end{eqnarray}

Due to periodicity of the network, the partial currents obey the relationship
$I_k(E) = I_m(E_{k-m})$, and the boundary population $n_0$ is related to the
node potentials $n_k$ as $n_0(E_k) = n_k(E)$. This allows us to reduce the
integration over energy in Eqs.\ (\ref{noise_eh}) and (\ref{I1}) to an
elementary interval $0<E<eV$,
\begin{eqnarray}
&\displaystyle I = {1\over e} \int_0^{eV} dE\, J(E), \quad J(E) =
\sum_{k=-\infty}^\infty I_k,
\label{I2} \\
&\displaystyle S = {2\over R_N} \int_0^{eV} dE \sum_{k=- \infty}^\infty\left[
2n_k(1-n_k) + {1\over 3}(R_N I_k)^2\right].\label{S2}
\end{eqnarray}

The ``potentials'' of the nodes outside the gap, $|E_k|> \Delta$, are equal
to local-equilibrium values of the Fermi function, $n_k(E) = n_F(E_k)$ at $k
\geq N_+$, $k \leq -N_-$. The partial currents flowing between these nodes,
\begin{equation} \label{outgap}
I_k = \left[ n_F(E_{k-1}) -n_F(E_k)\right]/R_N, \;\; k > N_+, \; k \leq -N_-,
\end{equation}
are associated with thermally excited quasiparticles. The subgap currents may
be calculated by Ohm's law for the series of $N_+ + N_-$ subgap resistors,
\begin{equation} \label{subgap}
I_k = {n_- - n_+ \over (N_+ + N_-)R_N}, \;\; -N_- < k \leq N_+,
\end{equation}
where $n_\pm(E) = n_F(E_{\pm N_\pm})$. From Eqs.\ (\ref{outgap}) and
(\ref{subgap}) we obtain the current spectral density in Eq.\ (\ref{I2}) as
$J(E) = 1/R_N$, which results in Ohm's law, $V=IR_N$, for the electric
current through the junction. This conclusion is related to our disregarding
the proximity effect and the normal scattering at the interface. Actually,
both of these factors lead to the appearance of SGS and excess or deficit
currents in the $I$-$V$ characteristic, with the magnitude increasing along
with the interface barrier strength and the ratio $\xi_0/d$
\cite{Bezuglyi2000}.

The subgap populations can be found as the potentials of the nodes of the
subgap ``voltage divider'',
\begin{equation} \label{divider}
n_k =  n_- - (n_- - n_+){N_- +k\over N_+ + N_-}.
\end{equation}

By making use of Eqs.\ (\ref{S2})-(\ref{divider}), the net current noise can
be expressed through the sum of the thermal noise of quasiparticles outside
the gap and the subgap noise, $S= S_> + S_\Delta$, where
\begin{equation} \label{S>}
S_>\! = \!{4T\over 3R_N}\!\left\{\! 2\left[n_F(\Delta)\! +\! n_F(\Delta\! +\!
eV)\right]\! +\!\! \left[ {eV\over T}\! +\! \ln \!{n_F(\Delta\! +\! eV) \over
n_F(\Delta)} \right]\! \coth{eV \over 2T}\right\}\!\!,
\end{equation}
and
\begin{eqnarray} \label{SD}
&\displaystyle S_\Delta = {2\over 3R_N}\int_0^{eV} dE\, (N_+ +
N_-)\left[f_{+-}+f_{-+} +2(f_{++}+f_{--})\right], \nonumber \\
&\displaystyle f_{\alpha\beta} = n_\alpha(1-n_\beta).
\end{eqnarray}

\begin{figure}
\epsfxsize=9cm\centerline{\epsffile{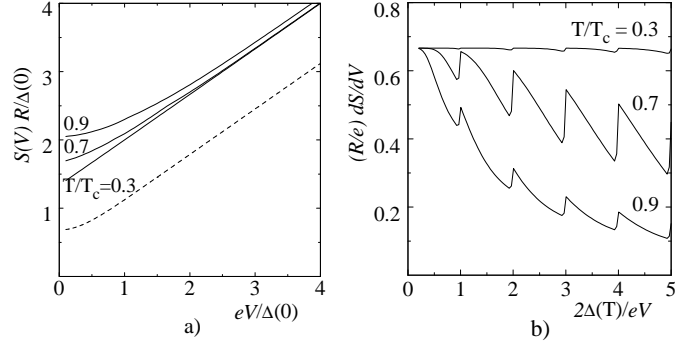}}
\caption{Spectral density $S$ of current noise vs voltage (a) and its
derivative $dS/dV$ vs inverse voltage (b) at different temperatures. Dashed
line shows the result for normal metal junction [10] at $T=0.3T_c$.}
\label{noisev}
\end{figure}

At low temperatures, $T \ll \Delta$, the thermal noise $S_>$ vanishes, and
the total noise coincides with the subgap shot noise, which takes the form
\begin{equation} \label{T=0}
S = {2\over 3R_N}\int_0^{eV} dE \, (N_+ + N_-) = {2\over 3R_N}(eV + 2\Delta),
\end{equation}
of $1/3$-suppressed Poisson noise $S=(2/3)q^{\it{eff}}I$ for the effective
charge $q^{\it{eff}} = e(1+ 2\Delta/eV)$ \cite{Bezuglyi2001}.
At $V \rightarrow 0$, the shot
noise turns to a constant value $4\Delta/3R$. At finite voltages, this
quantity plays the role of the ``excess'' noise, i.e. the voltage-independent
addition to the shot noise of a normal metal at low temperatures [see Fig.\
\ref{noisev}(a)]. Unlike short junctions, where the excess noise is
proportional to the excess current \cite{Hessling}, in our system the excess
current is small and has nothing to do with large excess noise.

Results of numerical calculation of the noise at finite temperature are shown
in Fig.\ \ref{noisev}. While the temperature increases, the noise approaches
its value for normal metal structures \cite{1/3Nagaev}, with additional
Johnson-Nyquist noise coming from thermal excitations. In this case, the
voltage-independent part of current noise may be qualitatively approximated
by the Nyquist formula $S(T) = 4 T^\ast/R$ with the effective temperature
$T^\ast = T + \Delta(T)/3$. The most remarkable phenomenon at nonzero
temperature is the appearance of steps in the voltage dependence of the
derivative $dS/dV$ at the gap subharmonics $eV = 2\Delta/n$ [Fig.\
\ref{noisev}(b)], which reflect discrete transitions between the
quasiparticle trajectories with different numbers of Andreev reflections. The
magnitude of SGS decreases both at $T \rightarrow 0$ and $T \rightarrow T_c$,
which resembles the behavior of SGS in the $I$-$V$ characteristic of long
ballistic SNS junction with perfect interfaces within the OTBK model
\cite{KBT 82,OTBK}. A small ``residual'' SGS in current noise, similar to the
one in the $I$-$V$ characteristic \cite{Bezuglyi2000}, should occur at $T
\rightarrow 0$ due to normal scattering at the interface or due to proximity
effect [see comments to Eq.\ (\ref{subgap})].

\subsection{Noise of tunnel barrier }

\begin{figure}
\epsfxsize=8.5cm\centerline{\epsffile{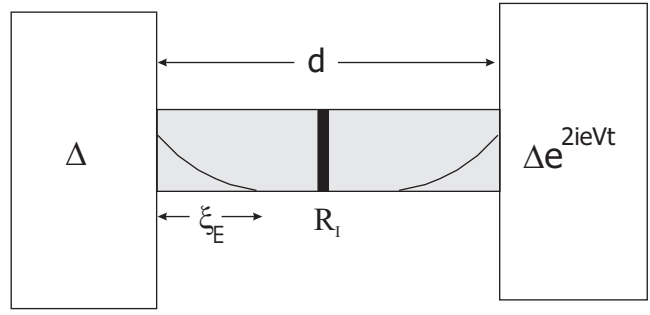}}
\caption{ Diffusive mesoscopic SNS junction with a tunnel barrier (bold
line).}\label{SNINS}
\end{figure}

It is instructive to compare the shot noise of a distributed source
considered in the previous section with the one produced by a localized
scatterer, e.g. opaque tunnel barrier with the resistance $R \gg R_N $
inserted in the normal region \cite{Bezuglyi 99}, see Fig. \ref{SNINS}. In
this case, the potential drops at the tunnel barrier, and the population
number is almost constant within the conducting region, while undergoes
discontinuity at the barrier. However, the recurrences in Eqs.
(\ref{boundary}), (\ref{recurr}) are not sensitive to the details of spatial
distribution of the population numbers, and therefore the result of the
previous section, Eq. (\ref{divider}), must be also valid for the present
case.  General equation for the noise in superconducting tunnel junctions has
been derived in Ref. \cite{LO2}. In our case of long SNINS junction, this
equation takes the form,
\begin{equation} \label{ShotNoise}
S =\int^{+\infty}_{-\infty}{dE\over R} \left[f(E)+f(E+eV)-f(E)f(E+eV)\right],
\end{equation}
where $f=n^e + n^h$. Taking into account the distribution function in Eq.\
(\ref{divider}), the noise power (\ref{ShotNoise}) at zero temperature
becomes
\begin{equation} \label{ShotNoise1}
S \!=\!{2\over R}\int^{-\Delta}_{-\Delta-eV}{dE\over3} \left[N_+ + N_-
+{2\over N_+ + N_-}\right].
\end{equation}

\begin{figure}
\epsfxsize=8.5cm\centerline{\epsffile{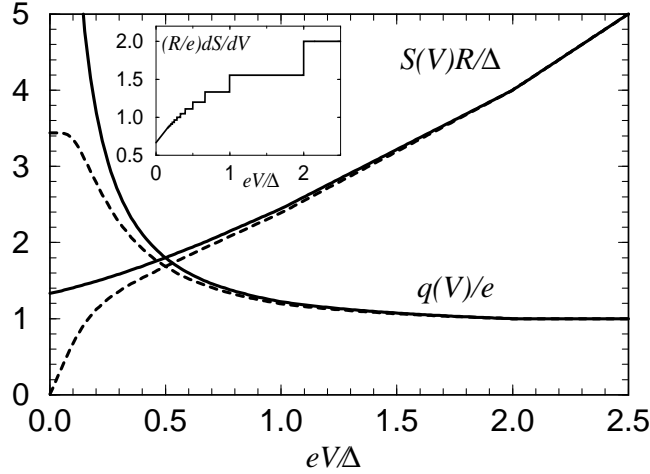}}
\caption{Spectral density $S(V)$ of current shot noise and effective
transferred charge $q(V)$ as functions of applied voltage $V$. In the absence
of inelastic collisions (solid lines), the shot noise power approaches the
finite value $S(0)=4\Delta/3R$ at $eV \rightarrow 0$, and the effective
charge increases as $q(V) = (e/3) (1 + 2\Delta/eV)$. The effect of inelastic
scattering is represented by dashed lines for the nonequilibrium parameter
$W_\varepsilon = 5$. The dependence $S(V)$ contains kinks at the gap
subharmonics, $eV = 2\Delta/n$, as shown in the inset. }\label{noiseSNINS}
\end{figure}

At voltages $eV>2\Delta$ this formula gives conventional Poissonian noise
$S=2eI$. At subgap voltages, the noise power undergoes enhancement: it shows
a piecewise linear voltage dependence,
$dS/dV=(2e/3R)[1+4/(\,\mbox{Int}\,(2\Delta/eV)+2)]$, with kinks at the
subharmonics of the superconducting gap, $eV_n=2\Delta/n$ (see Fig.
\ref{noiseSNINS}). At zero voltage, the noise power approaches the constant
value $S(0) = 4\Delta/3R$, coinciding with the noise power of the diffusive
normal region without tunnel barrier. However, in contrast to the latter
case, the voltage dependence of the noise here exhibits SGS already at zero
temperature, which consists of a step-wise increase of the effective charge
$q^{\it{eff}}(V)=S(V)/2I$ with decreasing voltage,
\begin{equation} \label{ChargeV}
{q^{\it{eff}}(n) \over e} = {1\over 3}\left(n+1+{2\over n+1}\right) =  1,
{11\over 9}, {22\over15}, \dots
\end{equation}
At $eV \rightarrow 0$ the effective charge increases as $q^{\it{eff}}(V)/e
\approx (1/3)(1+2\Delta/eV)$.

Calculation presented in this section shows that the $1/3$-factor in the
expression for the noise has no direct relation to the Fano factor for
diffusive normal conductors but is a general property of the incoherent MAR
regime. Appearance of this factor results from the infinitely increasing
chain of the normal resistances in the MAR network at small voltage, in close
analogy with the case of multibarrier tunnel structures considered in Ref.\
\cite{Jong}.

\section{Effect of inelastic relaxation on MAR}

During the previous discussion, an inelastic electron relaxation within the
normal region of the contact has been entirely ignored. This is legitimate
when inelastic mean free path exceeds the junction length, $l_\varepsilon =
\sqrt{{\cal D}\tau_\varepsilon} \gg d$, or, equivalently, when the inelastic
relaxation time exceeds the diffusion time, $\tau_\varepsilon \gg
\hbar/E_{\it Th}$.
\footnote{For the tunnel barrier, this estimate changes to $\tau_\varepsilon
\gg  (R/R_N)(\hbar/E_{\it Th})$ \cite{Bezuglyi 99}.}
In the opposite limiting case, $\tau_\varepsilon \ll  \hbar/E_{\it Th}$, the
energy relaxation completely destroys the MAR regime because the
quasiparticle distribution within the normal region becomes equilibrium, and
the SNS junction becomes equivalent to the two NS junctions connected in
series through the equilibrium normal reservoir. Thus the appearance of the
MAR regime is controlled by parameter $W_\varepsilon= E_{\it
Th}\tau_\varepsilon/\hbar$. The relaxation time in these estimates must be
considered for the energy of order $\Delta$, $\tau_\varepsilon(\Delta)$,
since non-equilibrium electron population under the MAR regime develops
within the whole subgap energy interval. However, to build up such a
nonequilibrium population at small voltage, a long time, $\tau_{\it dwell}
=(\hbar/E_{Th})(2\Delta/eV)^2$, is actually needed, which exceeds the
diffusion time $\hbar/E_{\it Th}$ by squared number of Andreev reflections
(see below Eq. (\ref{diffE}) and further text). Thus, the condition for the
MAR regime, $\tau_\varepsilon >
\tau_{\it dwell}$, is more restrictive than $ W_\varepsilon>1$, and in the
limit of zero voltage the MAR regime is always suppressed. Hence the effect
of the shot noise enhancement disappears at small voltage, and the noise
approaches the thermal noise level, the noise temperature being equal to the
physical temperature $T$ if the inelastic scattering is dominated by
electron-phonon interaction (assuming that the phonons are in equilibrium
with the electron reservoir). If the electron-electron scattering dominates,
the noise temperature may exceed the temperature $T$ of the electron
reservoir if this temperature is small, $T\ll\Delta$ (hot electron regime).
The reason is that at low temperature, the subgap electrons are well
decoupled from the reservoir (electrons outside the gap) due to weak energy
flow through the gap edges. In the cross over region of a small applied
voltage, the non-equilibrium electron population appears as the result of the
two competing mechanisms: the spectral flow upward in energy space driven by
MAR and spectral counter flow due to inelastic relaxation. In this section we
briefly consider the behavior of the shot noise in this situation.

To include the inelastic relaxation into the consideration we add the
inelastic collision term $I_{\varepsilon}$ to the diffusion equation
(\ref{DiffEq}),
\begin{equation} \label{diff}
{\cal D}{\partial^2 n \over \partial x^2} = I_{\varepsilon}(n).
\end{equation}
At small voltage, $eV\ll \Delta$, the spatial variation of the population
number is small, and the function $n(x)$ in the collision term can be
replaced by the boundary values, $n_0^{e,h}(E) \approx n_d^{e,h}(E) \equiv
n(E)$.  Then including the collision term into the recurrences, Eq.
(\ref{recurr}), and combining them with equations (\ref{boundary}), we arrive
at the generalized recurrence,
\begin{equation} \label{RecurrencesW}
{\cal D}\left[n(E + 2eV) + n(E - 2eV) - 2n(E)\right]= 4I_\varepsilon[n(E)].
\end{equation}
Within the same approximation, this recurrence is to be considered as
differential relation, which results in the diffusion equation for $n(E)$,
\begin{equation} \label{diffE}
D_E{\partial^2 n \over \partial E^2 } = I_{\varepsilon}(n),
\end{equation}
where $D_E=(eV)^2 E_{{Th}}/\hbar$ is the diffusion coefficient in energy
space.
\footnote{The finite resistance $R_{NS}$ of the NS interfaces, which
partially blocks quasiparticle diffusion, can be taken into consideration by
renormalization of the diffusion coefficient, $D_E \rightarrow D_E
[1+(d/\xi_0)(R_{NS}/R)^2]^{-1}$ (see Ref.\ \cite{Bezuglyi2000}).}

To demonstrate the effect of electron-phonon scattering and suppression of
the current shot noise, it is sufficient to assume the relaxation time
approximation in the collision term, $I_{\varepsilon}(n)=(1/\tau_\varepsilon)
[n-n_F(T)]$. The results of numerical calculations within this approximation
of the shot noise at zero reservoir temperature are presented in Fig.
\ref{noiseSNINS} by dashed curves. The rapid decrease of $S(V)$ at low
voltage is described by the following analytical approximation,
\begin{equation} \label{NoiseInelastic}
S(V) = S(0){3\over \alpha}\left( \tanh{\alpha\over 2} + {\alpha-\sinh\alpha
\over \sinh^2\alpha} \right), \quad \alpha = {\Delta\over
eV\sqrt{W_\varepsilon}},
\end{equation}
and it occurs when the length of the MAR path in energy space interrupted by
inelastic scattering, $eV\sqrt{W_\varepsilon}$, becomes smaller than
$2\Delta$.

\subsection{Hot electron regime}
In the case of dominant electron-electron scattering, equation (\ref{diffE})
describes the crossover from the ``collisionless'' MAR regime to the hot
electron regime as function of the parameter $D_E\tau_{ee}(\Delta)/\Delta^2$.
In the hot electron limit, $\Delta^2 \gg D_E\tau_{ee}$, the collision
integral dominates in Eq.\ (\ref{diffE}), and therefore the approximate
solution of the diffusion equation is the Fermi function with a certain
effective temperature $T_0 \ll \Delta$. The value of $T_0$ can be found from
Eq.\ (\ref{diffE}) integrated over energy within the interval $(-\Delta,
\Delta)$ with the weight $E$, taking into account the boundary conditions
$n(\pm \Delta) = n_F(\pm \Delta/T)$ and the conservation of energy by the
collision integral,
\begin{equation} \label{hot1}
D_E \left[1 -2\left(e^{-\Delta/T}+ {\Delta\over T_0}
e^{-\Delta/T_0}\right)\right] + 2 \int_\Delta^\infty dE\, E \,I_\varepsilon =
0.
\end{equation}

At zero temperature of the reservoir and $T_0 \ll \Delta$, Eq.\ (\ref{hot1})
takes the form
\begin{equation} \label{hot2}
D_E \tau_{\it ee} \Delta^2 = 2 \int_\Delta^{\infty} dE\, E
\int_{E-2\Delta}^\infty dE^\prime \int_{E-E^\prime-\Delta}^\infty d\omega\,
n_F(E-E^\prime- \omega)n_F(E^\prime) n_F(\omega),
\end{equation}
and can be reduced to an asymptotic equation for $T_0$,
\begin{equation} \label{asymptT}
(eV)^2 W_\varepsilon \exp(\Delta/T_0) = T_0 \Delta (1+ T_0/ \Delta),
\end{equation}
which shows that the effective temperature of the subgap electrons decreases
logarithmically with decreasing voltage. The noise of the hot subgap
electrons is given by the Nyquist formula with temperature $T_0$,
\begin{equation} \label{asymptS}
S(V) = (4T_0/ R)\left[1 - 2 \exp(-\Delta/T_0)\right],
\end{equation}
where the last term is due to the finite energy interval available for the
hot electrons, $|E|<\Delta$. Equations (\ref{asymptT}), (\ref{asymptS}) give
a reasonably good approximation to the result of the numerical solution of
Eq.\ (\ref{diffE}) \cite{NagaevSNS}.

\section{Summary}

We have presented a theory for the current shot noise in long diffusive SNS
structures with low-resistive interfaces at arbitrary temperatures. In such
structures, the noise is mostly generated by normal electron scattering in
the N-region. Whereas the $I$-$V$ characteristics are approximately described
by Ohm's law, the current noise reveals all characteristic features of the
MAR regime: ``giant'' enhancement at low voltages, pronounced SGS, and excess
noise at large voltages. The most spectacular feature of the noise in the
incoherent MAR regime is a universal finite noise level at zero voltage and
at zero temperature, $S= 4\Delta/3R$. This effect can be understood as the
result of the enhancement of the effective charge of the carriers, $q^{\it
eff}=2\Delta/V$, or, alternatively, as the effect of strongly non-equilibrium
quasiparticle population in the energy gap region with the effective
temperature $T_0=\Delta/3$. Appearance of the excess noise is controlled by a
large value of the parameter $\tau_{\varepsilon}/\tau_{\it dwell}\gg 1 $,
where $\tau_\varepsilon$ is the inelastic relaxation time, and $\tau_{\it
dwell}$ is the time spent by quasiparticle in the contact. At small applied
voltage this condition is always violated, and the excess noise disappears.
Under the condition of dominant electron-electron scattering, the junction
undergoes crossover to the hot electron regime, with the effective
temperature of the subgap electrons decreasing logarithmically with the
voltage. Calculation of the noise power has been done on the basis of circuit
theory of the incoherent MAR \cite{Bezuglyi2000}, which may be considered as
an extension of Nazarov's circuit theory \cite{CircuitTheory} to a system of
voltage biased superconducting terminals connected by normal wires in the
absence of supercurrent. The theory is valid as soon as the applied voltage
is much larger than the Thouless energy: in this case, the overlap of the
proximity regions near the NS interfaces is negligibly small, and the ac
Josephson effect is suppressed.


\end{document}